\documentclass[aps,superscriptaddress,twocolumn,showpacs]{revtex4}

\usepackage{graphicx}

\begin{document}

\title{Transition from Gaussian-orthogonal to Gaussian-unitary ensemble in a microwave billiard with threefold 
symmetry}

\author{R. Sch\"afer}
\affiliation{Fachbereich Physik, Philipps-Universit\"at Marburg, 
Renthof 5, 
D-35032 Marburg, Germany}
\author{M. Barth}
\affiliation{Fachbereich Physik, Philipps-Universit\"at Marburg, 
Renthof 5, 
D-35032 Marburg, Germany}
\author{F. Leyvraz}
\affiliation{Centro de Ciencias F\'{\i}sicas, Universidad Nacional 
Aut\'{o}noma de M\'{e}xico, Campus Morelos, C.~P. 62251, 
Cuernavaca, Morelos, M\'{e}xico}
\author{M. M\"uller}
\affiliation{Facultad de Ciencias, Universidad Aut\'{o}noma del 
Estado de Morelos, C.~P. 62210, Cuernavaca, Morelos, M\'{e}xico}
\author{T.H. Seligman}
\affiliation{Centro de Ciencias F\'{\i}sicas, Universidad Nacional 
Aut\'{o}noma de M\'{e}xico, Campus Morelos, C.~P. 62251, 
Cuernavaca, Morelos, M\'{e}xico}
\author{H.-J. St\"ockmann}
\affiliation{Fachbereich Physik, Philipps-Universit\"at Marburg, 
Renthof 5, 
D-35032 Marburg, Germany}

\date{\today}

\begin{abstract}
Recently it has been shown that time-reversal invariant systems 
with discrete symmetries may display in certain 
irreducible subspaces spectral statistics 
corresponding to the Gaussian unitary ensemble (GUE) rather than to 
the expected orthogonal one (GOE). 
A Kramers type degeneracy is predicted in such situations. 
We present results for a microwave billiard with a threefold 
rotational symmetry and with the option to display or 
break a reflection symmetry. This allows us to observe the change 
from GOE to GUE statistics for one subset of levels. 
Since it was not possible to separate the three subspectra reliably, 
the number variances for the superimposed spectra were studied.
The experimental results are compared with a theoretical and 
numerical study considering the effects of level splitting and level 
loss. 
\end{abstract}


\maketitle

\section{Introduction}

It is generally accepted that the eigenvalues of systems, whose 
classical 
counterparts are fully chaotic, behave similarly to those 
of random matrix ensembles \cite{Cas80,Boh84b}. This conjecture also 
has some theoretical foundations \cite{Ber85,Ley92,And96}. If we 
disregard spin there are two 
random matrix ensembles, with two different behaviors under time 
reversal. Systems with time reversal invariance (TRI) are 
described by an ensemble of real symmetric matrices, corresponding 
to the invariance of the Hamiltonian under complex conjugation. 
This ensemble is known as the Gaussian Orthogonal Ensemble (GOE) 
and should be contrasted with the ensemble of all Hermitian 
matrices, known as the Gaussian Unitary Ensemble (GUE), which 
usually applies when TRI is broken. 

An additional issue arises if discrete symmetries are present. In 
this case, it has been assumed that eigenvalues belonging to 
different irreducible representations of the symmetry group are 
statistically independent and obey GOE or GUE spectral statistics, 
depending on whether TRI holds or not. These assumptions have 
received considerable support both from theoretical considerations 
and from numerical simulations. However, in recent papers 
\cite{Ley96b,Ley96a}, it was suggested that an anomalous situation 
arises when a time-reversal invariant system has a discrete 
symmetry, such that TRI is broken within one of the irreducible 
subspaces of the point group. This is equivalent to the fact that 
the symmetry group has non-selfconjugate representations. In this 
case, TRI induces a degeneracy between different (conjugate) 
irreducible representations of the symmetry group (Kramers 
degeneracy). It was suggested that the eigenvalues belonging to 
these degenerate subspaces should have the statistical properties 
of the GUE rather than the GOE. This might be expected from the 
fact that complex conjugation actually maps one irreducible 
representation onto the other, rather than leaving it invariant. 
In this sense, one may indeed argue that TRI is violated within 
the subspace belonging to either representation, and is only 
restored through the combined presence of both representations. A 
more complete argument is presented in \cite{Ley96a}. A 
semi-classical view of this question is given in \cite{Kea97}. At 
this point we should add a note of caution. This entire argument 
applies to the largest symmetry group of the system, and relies on 
the fact that this group is known. Deviations from the 
expectations would indicate the presence of a larger unknown 
symmetry group. On the other hand, symmetry breaking, which is 
usually present in experiment, will also affect the results. In 
the present paper we will address both points in a simplest 
example. 

Let us now consider a system with threefold symmetry but with no 
twofold symmetry axis, i.\,e. $C_3$ symmetry. This means that the 
wave functions in polar coordinates can be put in the form 
\begin{equation}
\psi_{n,\sigma}(r,\phi)=\sum_{k=-\infty}^\infty\phi_{k,n}(r
)e^{i(3k+\sigma)\phi}\,,
\end{equation}
where $n$ runs over all the integers and $\sigma$ takes the values 
$0$ and $\pm1$. Because of TRI, it is clear that the eigenvalues 
corresponding to $\psi_{n,1}$ and $\psi_{n,-1}$ are degenerate, and 
we expect them to show GUE statistics. If in addition there is a 
twofold symmetry axis leading to $C_{3v}$ symmetry, the two conjugate 
representations combine to a two-dimensional irreducible 
representation which is self-conjugate. Thus we expect GOE 
statistics as well as degeneracy.

There is one previous microwave experiment on a billiard with 
$C_3$ symmetry performed by the Richter group \cite{Dem00b}. In 
the data analysis the authors faced the problem to separate the 
subspectra belonging to different irreducible representations. 
Since it is impossible to realize a $C_3$ symmetry exactly in 
experiment, the doubly degenerate eigenvalues belonging to the 
$\sigma=\pm 1$ representations are split into doublets. In 
Ref.\,\cite{Dem00b} this doublet splitting was used as a tool to 
separate the subspectra by classifying each pair of eigenvalues 
with a distance smaller than some given threshold value as a split 
doublet. The technique worked quite well, but there was a 
misidentification of about 5 percent of the levels.  
Nevertheless, the authors were able to see the GOE and GUE 
signatures in the respective subspectra. To improve the agreement 
with random matrix predictions, the authors calculated the 
spectrum numerically and used the results of the calculation to 
relabel the misidentified eigenvalues. 

In the present work a billiard with threefold symmetry is studied, 
which has the additional feature to allow a change of symmetry 
from $C_{3v}$ to $C_3$ by turning an insert placed in the center 
of the billiard, thus allowing to study the transition from GOE to 
GUE behavior of the $\sigma=\pm 1$ subspectra. 
In view of the problems to separate the respective subspectra 
satisfactorily, we refrained from any attempt in this direction. 
Instead we concentrated on the spectral 
statistics for the superimposed spectrum and compared the results 
with numerical and analytic calculations based on random matrix 
theory (RMT). 

A formula for the number variance is derived assuming 
that the degeneracy is lifted by a random off-diagonal matrix element 
between the two states of the doublets.  It describes fairly well 
the long range properties of the spectra completed by means of
level dynamics. 
The spectral short range behavior is sensitive to the details of the 
completion procedure while the long range behavior is not. This is 
why the difference between the GUE and the GOE case can still be
detected very reliably in the long range regime. 

If we consider the individual spectra there is a loss of roughly 
20\% of the levels. The number variance of the uncompleted spectra 
were compared with numerical RMT simulations. Several ways to 
simulate the loss of levels have been applied, and in all cases the 
two symmetries remained distinguishable at a loss rate of 
20\%. 

In the following section the experimental setup and the details of 
the data analysis are presented. Further the gradual transition from 
the $C_{3v}$ to the $C_3$ symmetry will be observed in the number 
variance of the experimental spectra.
In section III the results for the completed spectra will be compared 
with calculations taking into account the break of the $C_3$ 
symmetry. 
Finally, in section IV we will discuss the results for individual 
spectra with missing levels; these will be complemented with 
numerical RMT simulations. 

\section{Experimental technique}

\begin{figure}
\centerline{ \includegraphics[width=6.5cm]{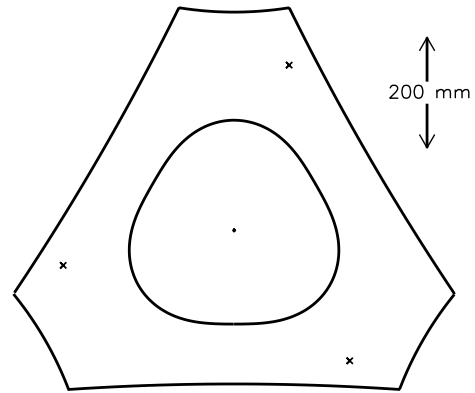}} 
\caption{Sketch of the microwave billiard used in the experiment 
at angle $\phi =0^\circ$. The positions of the antennas are marked by 
crosses.} 
 \label{fig:sketch}
\end{figure}

The main features of the experimental technique have been 
presented elsewhere \cite{Kuh00b}. Therefore we concentrate on the 
aspects of relevance in the present context. Figure \ref{fig:sketch} 
shows a sketch of the billiard used in the experiment. The 
interior insert can be rotated versus the outer part thus allowing 
a spectral level dynamics measurement in dependence of the 
rotation angle $\phi$. To avoid a break of symmetry due to the 
presence of the antenna, three antennas were placed at symmetry 
equivalent positions, two of them being terminated by $50\,\Omega$ 
loads. Spectra were taken for angles from 
$\phi=0^\circ$, corresponding to the geometry with $C_{3v}$ 
symmetry shown in the figure, to $\phi=20^\circ$ in steps of one 
degree. All boundaries, both of the insert and the outer part, 
were curved, and the corners of the outer boundary were removed to 
avoid families of marginally stable orbits such as bouncing balls. 
This was checked by means of a Poincar\'e plot of the classical 
motion of a particle in the billiard. 

\begin{figure}
 \centerline{ \includegraphics[width=7.5cm]{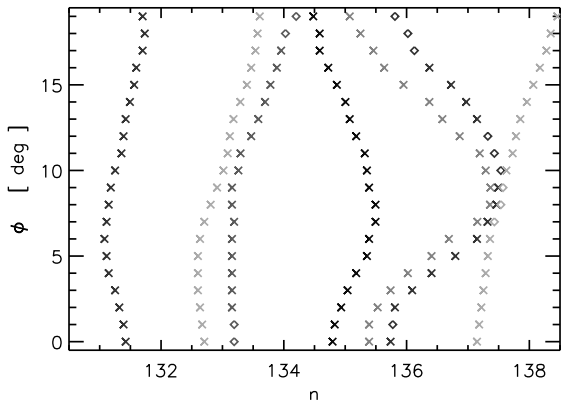} } 
\caption{Part of the spectral level dynamics for the billiard with 
threefold symmetry with the orientation angle $\phi$ of the 
central insert as the level dynamics parameter. Angle $\phi=0^\circ$ 
corresponds to $C_{3v}$ symmetry. Only the resonances marked by 
cross-shaped symbols were really found in the experiment. At 
the positions marked by diamond-shaped symbols missing resonances 
were supplemented.} 
 \label{fig:leveldyn} 
\end{figure}

The height of the billiard was $h=10$\, mm, i.\,e. the system was
quasi-two-dimensional for frequencies below $\nu_{\rm 
max}=c/2h=15$\,GHz. Microwave reflection spectra were taken with a 
Wiltron 360B vector network analyzer in the frequency range 0.5 to 
7 GHz with a resolution of 0.5 MHz. 
In the studied spectral range one expects to find altogether 400 
eigenvalues for each angle $\phi$. This number is obtained  
from the Weyl formula with area and circumference term
\begin{equation}\label{1}
  \bar{n}(k)=\frac{A}{4\pi}k^2-\frac{S}{4\pi}k\,.
\end{equation}                                    
$\bar{n}(k)$  is the mean integrated density of states, where $A 
\approx 0.258$\,m$^2$ and $S \approx 3.59$\,m are area and 
circumference of the billiard, and $k=2\pi\nu/c$ is the wave 
number. The curvature of the billiard yields a correction term of 
the order of 1. The number of states identified for a fixed angle 
falls 
approximately 20\% short of the Weyl estimate. It is possible to 
recover all missing eigenvalues by studying the dynamics of the 
spectra as a function of the angle. This is demonstrated in 
Fig.\,\ref{fig:leveldyn} where the positions of the missing 
resonances are 
marked by diamond-shaped symbols. 

We begin with a presentation of the number variance in dependence of 
the insert angle, both for the completed and uncompleted spectra.
In both cases the spectra were unfolded to a constant mean density 
of states using a second order polynomial fit to the 
experimental integrated density of states. 
The number variance $\Sigma^2(L)$ has been obtained by moving the 
interval of length $L$ through each spectrum on a fine grid to not 
loose any information. Consequently, the smoothness of $\Sigma^2(L)$ 
does not reflect its statistical quality, since the intervals are not 
statistically independent.

\begin{figure}
 \centerline{ \includegraphics[width=7.5cm]{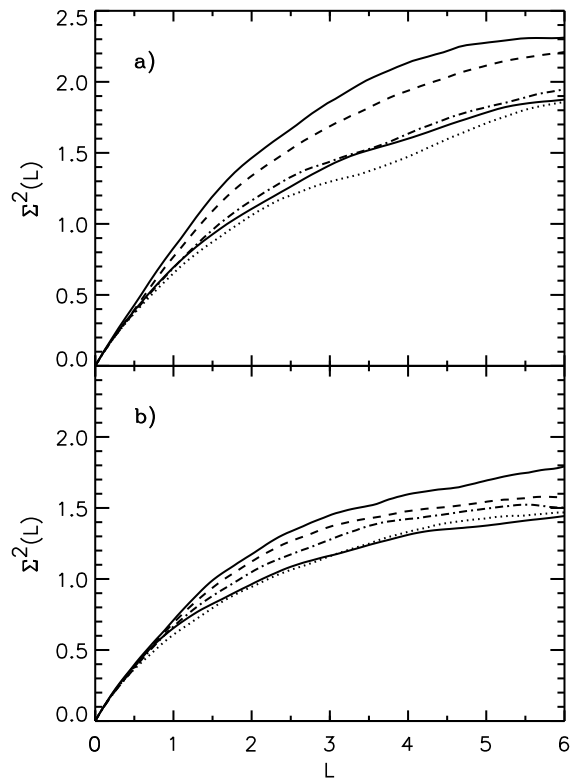}} 
\caption{Number variance $\Sigma^2(L)$ for different angles $\phi$ 
for the completed (a) and the uncompleted (b) spectra. 
The results for $\phi=0^\circ, 1^\circ$ (upper solid line), $\phi=2^\circ, 
3^\circ$ (dashed), $\phi=4^\circ, 5^\circ$ (dash-dotted), $\phi=6^\circ, 7^\circ$ (dotted), were averaged. The lower solid line shows an average over 
$\phi=8^\circ\ldots 19^\circ$.}
 \label{fig:s2_trans} 
\end{figure}

Figure \ref{fig:s2_trans} shows the number variance $\Sigma^2(L)$ for 
different angles $\phi$, where the results for $\phi=0^\circ, 
1^\circ$, $\phi=2^\circ, 3^\circ$, $\phi=4^\circ, 5^\circ$ and 
$\phi=6^\circ, 7^\circ$ have been averaged. 
We see a gradual transition from the $C_{3v}$ to 
the $C_3$ case with increasing $\phi$. The line with long dashes shows an 
average over $\phi=8^\circ\ldots 19^\circ$, corresponding to the 
$C_3$ case.
The GOE-GUE transition is clearly seen, both in the completed and the 
uncompleted spectra.
In the following we will only discuss the two extreme cases, where 
for the $C_{3v}$ case the results for $\phi=0^\circ, 1^\circ$ will be 
averaged and for the $C_3$ case the ones for $\phi=8^\circ\ldots 
19^\circ$.

\section{Results for the completed spectra and the doublet splitting}

We recall that for $\phi=0^\circ$ we expect 
a superposition of one GOE spectrum and another doubly degenerate 
GOE spectrum for the ideal system, whereas for angles $\phi$ 
sufficiently different from zero a superposition of one GOE spectrum 
with a doubly degenerate GUE spectrum should be observed. 
The short range behavior is dominated by doublet splitting, which we 
have no reliable experimental handle on. We therefore concentrate on 
the long range behavior. 
In particular we consider the long range part of the two-point 
correlation function.
The number variance is ideally suited for this purpose because of the 
clear 
signature it shows for the difference between the GOE and the GUE 
case.

For the superposition of $M$ strictly non-interacting, equally 
weighted subspectra the number variance can be written as 
$$ 
\Sigma_{\rm{total}}^{2}(L)=\sum_{m=1}^{M}\Sigma_{m}^{2}
\left(\textstyle{\frac{1}{M}}L\right) \,.$$ 
In the case of a threefold symmetry, the three 
subspectra are not independent; instead two of them are degenerate 
and thus the covariance of this doublet spectrum leads to a factor 
2 in the number variance of their superposition: 
\begin{equation}
\Sigma_{\rm{total}}^{2}(L)=\Sigma_{1}^{2}
\left(\textstyle{\frac{1}{3}}L\right) 
+ 4\, \Sigma_{2}^{2}\left(\textstyle{\frac{1}{3}}L\right)  \,.
\label{s2total_deg}
\end{equation}
Here $\Sigma_1^2$ is the number variance of the singlet spectrum 
being a 
GOE spectrum in both cases, whereas $\Sigma_2^2$ is the number 
variance of 
the doublet spectrum which may be a GOE or a GUE spectrum, depending 
on 
the symmetry of the system.

In the analysis of the completed spectra, the first 70 eigenvalues 
of each spectrum have been omitted to avoid non-generic features 
in the statistics. The results for the number variance of the 
experimental spectra have been averaged  over the rotation angles 
$\phi=0,1^{\circ}$ in the $C_{3v}$ case and $\phi=8..19^{\circ}$ 
in the $C_3$ case. This implies that the statistical 
fluctuations of the number variance are much larger in the former 
case. 

\begin{figure}
\centerline{ \includegraphics[width=7.5cm]{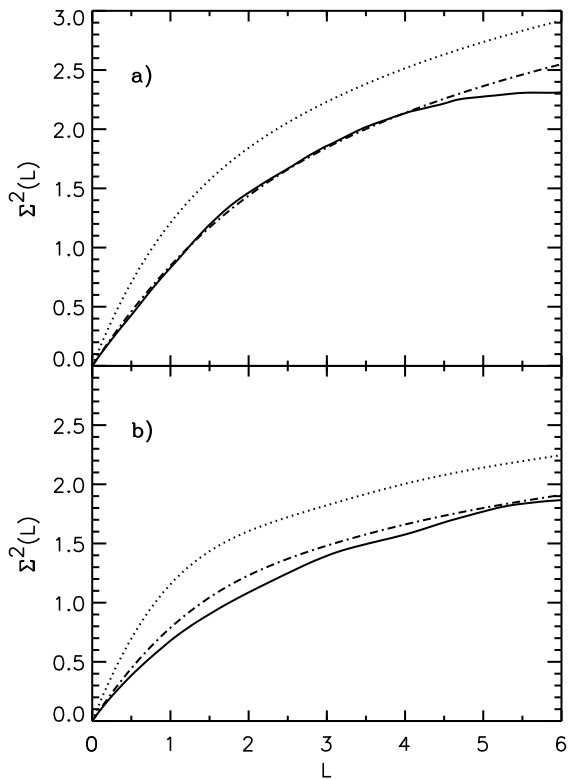}}
\caption{Number variance $\Sigma^2(L)$ of the completed spectra for 
the mirror-symmetric (a) and the non-mirror-symmetric (b) case. The 
experimental results (solid line) are compared with the theoretical 
prediction for the ideal system (dotted line), and with an analytical 
result, where the doublet splitting was taken into account 
(dash-dotted line).}
\label{fig:s2all}
\end{figure}

Figure \ref{fig:s2all} shows experimental data with solid lines for 
the $C_{3v}$ (a) and the $C_3$ (b) case, respectively, while the 
dotted lines give the ideal result of Eq.\,\ref{s2total_deg}. It is 
not surprising that the agreement is poor, since we 
already noticed in the level dynamics that the doublet spectrum 
was rather strongly split, indicating a sizeable break of the 
$C_3$ symmetry of the billiard. On the other hand, we clearly see 
that the two number variances differ, and that the effect is quite 
large. 
 This was the motivation for further theoretical 
considerations of the symmetry breaking. 
 A detailed discussion of this 
issue follows in the appendix, where an expression for the 
corresponding number variance is derived. Here only the main points 
shall be outlined.

The case of the $C_3$ symmetry is the more delicate one. The basis 
functions spanning 
the irreducible subspaces are complex, but due to the degeneracy 
we can always choose real functions. The appendix gives formal 
explanations, but here it 
may suffice to say that the actual measurements are on real 
fields and therefore it is clear that the smallest perturbation 
that does lift the degeneracy must cause real linear combinations. 
Therefore it is not surprising that we have to use degenerate 
perturbation theory for a real symmetric matrix. Furthermore it is 
quite clear that the perturbation matrix elements should be 
Gaussian distributed. While the mechanical imperfections causing the 
breakdown of the symmetry are the same for all functions, the 
unperturbed eigenfunctions involved are those of a chaotic system 
and thus in good approximation random functions that should yield 
Gaussian distributed matrix elements. Based on these simple 
considerations it is clear that the level splitting is Gaussian 
distributed, too,  
and linear in the perturbation. In the appendix we 
obtain an expression for the effect of such a level splitting on 
the long range part of the number variance, and we also show that 
effects of three level terms are small and thus effects of the 
symmetry breaking on the number variance for larger distances are 
of second order in the perturbation. 

This result is of central importance since it explains why the 
difference in the number variances persists in spite of the fairly 
large doublet splitting. A parameter $\Delta=0.125$ for the 
width of the distribution of the perturbation matrix element and 
thus for the doublet splitting (see Eq.\,\ref{def_distribution}), is 
consistent with the splitting 
found in the experiment. It corresponds to a mean level splitting of 
about 0.85 in units of the mean level spacing. 
If we now use the results of the appendix for the number variance, 
Eq.\,\ref{eq:total}, both for the GOE and the GUE case, we find good 
agreement with the experimental results, as can be inferred  from the 
dash-dotted line in Fig.\,\ref{fig:s2all}. The small deviations are 
consistent with the statistical uncertainties seen in skewness and 
excess (not shown). We therefore have a clear understanding of the 
role played by the splitting of the degeneracy and why it does not 
affect the difference between the two cases even if the symmetries 
are only approximate.

\section{Missing levels and individual spectra}

Completing the spectra by means of spectral level dynamics we were
able to come to a quantitative theoretical understanding of the 
GOE-GUE transition observed in the experiment. 
The completion of spectra is definitely a legitimate approach, but 
in general only isolated spectra for fixed systems are available. 
Therefore we felt the need to check the robustness of the signatures 
of the transition by considering what we can learn from incomplete 
spectra with as many as 20\% of the levels missing. 

Numerical simulations have been performed as follows. We 
produced ensembles of 1000 spectra of dimension 500 to create the 
GOE or the GUE doublets for the $C_{3v}$ and $C_3$ case 
respectively. Then we lifted the degeneracy randomly using a 
Gaussian distribution with a width of about 12.5\% of the average 
level distance of a single spectrum. These spectra have been 
superposed with an independent GOE spectrum of the same strength. 

In the next step we tried to simulate the experimental loss.
There are two sources of missing levels. First, close pairs can no 
longer be resolved if their separation is smaller than the line width.
Second, a level will be missed if incidentally a node line of 
the associated wave function will be just at the position of the 
antenna.
This mechanism leads to a loss that is uniformly distributed along 
the energy axis.

So we skipped randomly 7\% of the eigenvalues to simulate 
the global loss due to the position of the antenna. Then we 
compared the distance between two adjacent levels with a Gaussian 
distributed random number whose width is chosen such that around 
13\% of the doublets are destroyed. From these spectra only 30\% 
of the levels at the center of each spectrum are taken into 
account for the further analysis in order to avoid edge effects. 
The ratio of uncorrelated and correlated loss has been chosen to 
get a fair correspondence between experiment and theory for 
$\Sigma^2$ at small distances and is consistent with values found 
in previous works (see e.\,g. Ref.\,\cite{shu94}). 

\begin{figure}
 \centerline{ \includegraphics[width=7.5cm]{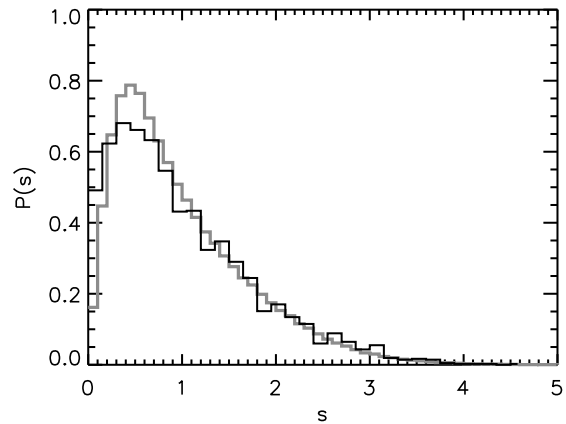}} 
\caption{Nearest neighbor distribution for the case of $C_3$ 
symmetry. The black line shows the results for the experimental 
data with 20\% loss, the grey line shows the numerical 
simulations.} 
 \label{fig:nndloss} 
\end{figure}

\begin{figure}
 \centerline{ \includegraphics[width=7.5cm]{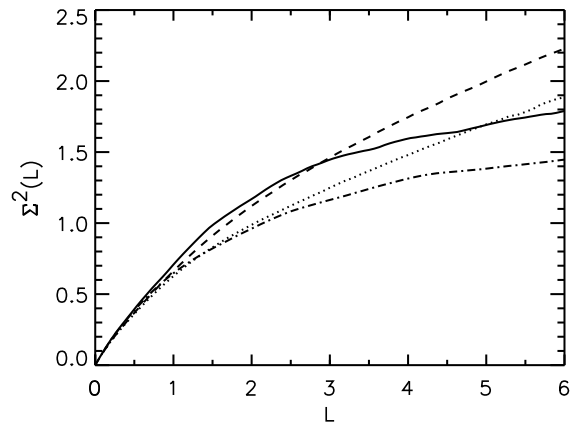}} 
\caption{Number variance $\Sigma^2(L)$ of the experimental data 
with 20\% loss (solid line: $C_{3v}$, dash-dotted line: 
$C_3$ case) in comparison with the numerical 
simulations (dashed line: $C_{3v}$, dotted line: $C_3$ case).
}
 \label{fig:s2loss} 
\end{figure}

The numerical as well as the experimental results for the nearest 
neighbor distribution for the case of $C_3$ symmetry are presented 
in Fig.\,\ref{fig:nndloss}. A good qualitative agreement
between the two curves is found, although the first bins are 
underestimated 
in the numerical data resulting in an overshoot of its maximum. 
In Figure \ref{fig:s2loss} the experimental results for 
both symmetries are  compared with the numerics. In both cases we 
get a rough agreement up to $L=3$. For larger 
distances the slope of the numerical curve is by far too large which 
indicates that we have somehow destroyed long range correlations 
still present in the experimental data. On the other hand, the 
reliability of the experimental data for large $L$ is 
questionable because of the missing levels. Therefore it 
seems  unreasonable to fit parameters like the percentages of the 
two sources of level missing. In any case, both the experimental 
results as well as the numerical calculations permit a clear 
distinction between the two symmetry cases. 

\section{Conclusions}

We have presented experimental results for a quasi-two-dimensional 
microwave cavity with a movable insert
permitting to change the discrete symmetry of the cavity from 
$C_{3v}$ to $C_3$. According to theory, this change should modify the 
spectral statistics in the ideal case from a superposition of 
two GOE spectra, one of them degenerate, to a superposition of a 
GOE spectrum with a degenerate GUE spectrum. 
We find that the level splitting due to imperfections in the 
$C_3$ symmetry amounts to about 0.85 of the mean level spacing, 
and we face up to 20\% loss of levels in an isolated spectrum.
From spectral level dynamics by rotating the cavity's insert we can 
recover the 
entire spectra, with some errors in the short range behavior of 
correlation functions. Nevertheless, the long range part of the 
number variance still clearly displays the difference between the 
two cases. From a theoretical analysis of the symmetry breaking in 
terms of degenerate perturbation theory, this behavior becomes 
comprehensible, since the level splitting is linear in the 
perturbation while long-range effects are quadratic. 

Even more importantly, we have shown that the spectral statistics of spectra 
with large loss of levels can still provide usable information.
By postulating two types of losses, a statistical one due to the 
fact that the antenna may be close to a node of the 
wavefunction and a systematic one due to overlap of close-lying 
levels, we can explain the experimental results qualitatively.

\begin{acknowledgments}
The experiments were supported by the Deutsche 
Forschungsgemeinschaft. Travel support was obtained from CONAyT 
32101-E and 32173-E as well as from DGAPA IN112200 projects. 
Discussions took place at workshops at the Centro Internacional de 
Ciencias in Cuernavaca, Mexico, in 1998, 2000 and 2001. M.\,B., R.\,S.\ and 
H.-J.\,S.\ would like to thank the organizers and the center for the 
hospitality and inspiring atmosphere.
T.H.\,S.\ acknowledges support of the A.-v.-Humboldt foundation. 
\end{acknowledgments}
 
\appendix

\section{Number Variance for perturbed $C_3$ symmetry} 
\label{sec:s2break}

In this appendix we present a detailed treatment of the effect of 
symmetry breaking. We first need to specify a reasonable RMT model 
for this system. To this end let us consider an unperturbed 
Hamiltonian matrix $H_0$ representing the billiard with perfect 
$C_3$ symmetry. The Hilbert space can be separated into three 
irreducible parts, one invariant under the group action, the other 
two spanned by clockwise and counterclockwise travelling waves. 
Because of TRI the eigenfunctions in the latter two subspaces have 
pairwise degenerate eigenvalues, i.\,e. they display a Kramers 
degeneracy. If we wish to display TRI rather than the $C_3$ 
invariance we can alternatively use a real basis for the doublet 
space. Any perturbation due to imperfections of the billiard can 
only break the $C_3$ symmetry but not TRI. As we wish to introduce 
an RMT model for the perturbation that conserves TRI, it is 
obvious that this can be implemented by a real symmetric matrix in 
the real basis.

From this picture the effect of the perturbation is 
immediately clear: Since we have degenerate eigenvalues, we must 
look at these separately. Then we may 
ask what happens if there is a third eigenvalue in the 
vicinity. We therefore consider the matrix 
\begin{eqnarray}
A=\left( \matrix{ 0&V_1&V_2\cr V_1&0&V_3\cr
V_2&V_3&1\cr } \right), \label{eq:matrix}
\end{eqnarray}
where we have scaled and shifted the two degenerate eigenvalues to 
zero, and the perturbing third eigenvalue one. The eigenvalues 
$\lambda_i$ obey the relations
\begin{eqnarray}
\lambda_1+\lambda_2+\lambda_3&=&1\nonumber\\
\lambda_1^2+\lambda_2^2+\lambda_3^2&=&1+2(V_1^2
+V_2^2+V_3^2) \label{eq:eigenvalue}\\
\lambda_1\lambda_2\lambda_3&=&-V_1^2+2V_1V_2
V_3\,.\nonumber
\end{eqnarray}
These equations are satisfied to second order in the $V_i$ by
\begin{eqnarray}
\lambda_1&=&V_1\nonumber\\ \lambda_2&=&-V_1
\label{eq:solution}\\ \lambda_3&=&1\,.\nonumber
\end{eqnarray}
The shift in the two degenerate
eigenvalues, being linear in the off-diagonal
element $V_1$, thus dominates all other perturbations as long as the 
$V_i$ are small with respect to the third eigenvalue. If on the
other hand the third eigenvalue is so close to the
degenerate doublet that the $V_i$ are of the order of one,
the situation becomes more complex and Cardan's formula must
be used. However, the splitting will still be linear in the $V_i$, so 
that the qualitative behavior should 
not be greatly affected. The probability for one doublet to come 
close to another one is very small, since they experience GUE level 
repulsion. 

We now turn to the calculation of the spectral form factor, and of 
the number variance for a weakly split GUE doublet spectrum. Let us 
take 
the $\epsilon_k$ to be an arbitrary reference spectrum with mean 
level spacing equal to one. The split doublet spectrum is generated by
\begin{equation}
E_{k,\sigma}=2\epsilon_k+\sigma a_k\,, \label{eq:def_spectrum}
\end{equation}
where $\sigma$ runs over $\pm1$ and the factor 2 retains the mean 
level spacing of one. The $a_k$ are
taken to be independent Gaussian random variables distributed
according to
\begin{equation}
p(a)=\sqrt{\frac{1}{2\pi\Delta}}\exp\left(-a^2/(2\Delta)\right).
\label{def_distribution}
\end{equation}
We are now going to compute the form factor
\newcommand\sigmap{\sigma^\prime}
\begin{equation}
k(t)=\frac{1}{2N}\sum_{\sigma,\sigmap=\pm1} 
\sum_{k,l=1}^{N}e^{2\pi  i(E_{k,\sigma}-E_{l,\sigmap})t} 
\label{eq:def_k} 
\end{equation}
from the corresponding quantity $k_0(t)$ for the $\epsilon_k$. 
Separating the right hand side into a part for $k=l$ and another for 
$k\neq l$ and averaging over the $a_k$ yields 
\begin{equation}
k(t)=1+e^{-8\pi^2 \Delta t^2}-2 e^{-4\pi^2 \Delta 
t^2}+2k_0(2t)e^{-4\pi^2 \Delta t^2}\,. \label{eq:final} 
\end{equation}
The number variance $\Sigma^2(L)$ can be expressed as (see e.\,g. 
Ref.\,\cite{Ber88})
\begin{equation}
\Sigma^2(L)=\frac{2}{\pi^{2}}\int_{0}^\infty 
dt\,k(t)\left(\frac{\sin \pi Lt}{t}\right)^2\,. \label{eq:sigma2} 
\end{equation}
Thus, from Eq.\,\ref{eq:sigma2} we get the $\Sigma^2_d(L)$ for the 
doublet part
of the spectrum by inserting the GUE form factor
\begin{eqnarray}
k_0(t)&=&|t|\qquad(|t|\leq1) \label{eq:GUE}\\
&=&1\qquad(|t|>1)\nonumber
\end{eqnarray}
into (\ref{eq:final}). From this one finally obtains for the number 
variance of the superimposed spectrum 
\begin{equation}
\Sigma^2_{total}(L)=\Sigma^2_{d}\left( \textstyle{\frac{2}{3}}L\right)
+\Sigma^2_{1}\left( \textstyle{\frac{1}{3}}L\right)\,, 
\label{eq:total} 
\end{equation}
where $\Sigma^2_{1}(L)$ is the number variance of the GOE singlet 
spectrum.

The $C_{3v}$ case yields a similar result if we 
substitute the GOE form factor rather than the GUE form factor for 
$k_0$ in Eq.\,\ref{eq:final}.


\begin{thebibliography}{10}

\bibitem{Cas80}
G. Casati, F. Valz-Gris, and I. Guarneri, Lett. Nuov. Cim. {\bf 28},  
279
  (1980).

\bibitem{Boh84b}
O. Bohigas, M. Giannoni, and C. Schmit, Phys. Rev. Lett. {\bf 52},  
1  (1984).

\bibitem{Ber85}
M. Berry, Proc. R. Soc. Lond. A {\bf 400},  229  (1985).

\bibitem{Ley92}
F. Leyvraz and T. Seligman, Phys. Lett. A {\bf 168},  348  (1992).

\bibitem{And96}
A. Andreev, O. Agam, B. Simons, and B. Altshuler, Phys. Rev. Lett. 
{\bf 76},
  3947  (1996).

\bibitem{Ley96b}
F. Leyvraz and T. Seligman,  in {\em Proceedings of the IV 
Wigner-Symposium in
  Guadalajara, Mexico}, edited by N. Atakishiyev, T. Seligman, and K. 
Wolf
  (World Scientific, Singapore, 1996), p.\ 350.

\bibitem{Ley96a}
F. Leyvraz, C. Schmit, and T. Seligman, J. Phys. A {\bf 29},  L575  
(1996).

\bibitem{Kea97}
J. Keating and J. Robbins, J. Phys. A {\bf 30},  L177  (1997).

\bibitem{Dem00b}
C. Dembowski {\it et~al.}, Phys. Rev. E {\bf 62},  R4516  (2000).

\bibitem{Kuh00b}
U. Kuhl, E. Persson, M. Barth, and H.-J. St\"ockmann, Eur. Phys.~J. B 
{\bf 17},
   253  (2000).

\bibitem{shu94}
A. Shudo {\it et~al.}, Phys. Rev. E {\bf 49},  3748  (1994).

\bibitem{Ber88}
M. Berry, Nonlinearity {\bf 1},  399  (1988).

\end{thebibliography}

 \end{document}